\documentclass[final,1p,12pt]{elsarticle}
\pdfoutput=1
\usepackage[hyphens]{url}
\usepackage{hyperref}
\usepackage[hyphenbreaks]{breakurl}

\usepackage{amssymb}
\usepackage{dirtytalk}

\usepackage{listings}
\usepackage{color}

\definecolor{dkgreen}{rgb}{0,0.6,0}
\definecolor{gray}{rgb}{0.5,0.5,0.5}
\definecolor{mauve}{rgb}{0.58,0,0.82}

\makeatletter
\def\ps@pprintTitle{%
 \let\@oddhead\@empty
 \let\@evenhead\@empty
 \def\@oddfoot{\centerline{\thepage}}%
 \let\@evenfoot\@oddfoot}
\makeatother

  \definecolor{light-gray}{gray}{0.80}

\journal{}
\begin{document}

\begin{frontmatter}



\title{Emergency Management and Recovery of Luna Classic}


\author{Edward Kim}
\author{Tobias Andersen}
\author{Marventus}
\author{A.E.}
\author{Pedro Borges}
\author{David Schmidt}
\author{Matthew Western}
\affiliation{organization={ek826, Zaradar, Marventus, A.E., Vegas, Resolvance, Raider70},
                country={[on Discord]}}
\begin{abstract}
In early May 2022, the Terra ecosystem collapsed after the algorithmic stablecoin failed to maintain its peg.  Emergency measures were taken by Terraform Labs (TFL) in an attempt to protect Luna and UST, but then were abruptly abandoned by TFL for Luna 2.0 several days later.  At this time, the Luna Classic blockchain has been left crippled and in limbo for the last two months.  

In the face of impossible odds, the Luna Classic community has self organized and rallied to build and restore the blockchain.  This technical document outlines the steps we, the community, have taken towards the emergency management of the Luna Classic blockchain in the weeks after the UST depeg.   We outline precisely what would be implemented on-chain to mitigate the concerns of affected stakeholders, and build trust for external partners, exchanges, and third-party developers.  For the Luna Classic community, validators, and developers, this outlines concrete steps on how passed governance can and will be achieved. We openly audit our own code and welcome any feedback for improvement.  Let us move forward together as the true community blockchain.  
\end{abstract}

\end{frontmatter}


\newpage
\tableofcontents
\newpage
\section{Introduction}
\label{sec:intro}
The Terra framework consisted of Terra LUNA and Terra UST.  This pair represented an algorithmic stablecoin that pegged UST to 1 USD using a swap and mint function between LUNA and UST. When UST’s value dropped below one dollar, UST asset holders could burn their UST, minting one Luna thus decreasing UST in circulation, and increasing its price. When UST’s value increased over one dollar, LUNA asset holders could burn one LUNA and mint one UST giving a small profit and lowering the price of UST back to 1 USD. 

However, on May 10th 2022, UST depegged from its one-dollar mark and started a forty percent sell-off within 24 hours. In the span of four days, the price of UST crashed to only thirteen percent from its one-dollar stablecoin value.  Luna Foundation Guard deployed over \$3 billion in BTC to defend the peg.  Further, as designed, the algorithm that linked UST and LUNA minted trillions of LUNA to avoid the collapse of UST, but the framework could not support the massive sell-off and the Terra ecosystem suffered a devastating death spiral.

\subsection{Emergency Measures during UST Depeg}
During the collapse of the stablecoin, Terraform Labs initiated three emergency actions.  The first was to burn the UST in the community pool, the second was to burn the 371 million UST cross-chain on Ethereum, and the third was to stake 240 million LUNA to defend from governance attacks\footnote{\url{https://twitter.com/terra\_money/status/1524654729753956352?s=20&t=WFQ7l1rYHM4bN4uwl1Mokg}}. However, user ``Dev'' (@valardragon) noted on Twitter that the LUNA chain was in danger and could be taken over for 4M at that moment\footnote{\url{https://twitter.com/valardragon/status/1524732164784496640}}.  A patch was devised that disabled IBC channels, delegation, and creating new validators.   The TFL developer who created the security patch, (@alexanderbez), stated that ``Once... what is being termed as Terra Classic launches, this will be reverted obviously.''   The swap connection between LUNA and UST was also disabled.  One of the primary reasons for this was due to numeric precision, i.e. the market module was running out of decimal places for the supply.  See our postmortem of the depeg on our GitHub \cite{postmortem1}.

\subsection{Governance and Representation}
While the TFL security patch halted the hyperinflation of LUNA and protected the chain from attack, the patch was a change to the protocol, which according to the Terra Classic documentation and governance procedures\cite{governance} requires a proposal and community vote, but no proposal was published and a vote did not happen, and so the patch was installed in contravention to these accepted governance procedures. That being said, we have no reason to doubt that the developers and validators acted in good faith and made an executive decision in times of emergency.  

\subsection{Stewardship by Terraform Labs}
Since that moment two months ago, there has been no word from TFL on the support or continuance of support for the Luna Classic blockchain.  Numerous attempts to contact Terraform Labs (TFL) have resulted in frustrating silence.  Attempts from the authors to communicate with TFL include,
\begin{enumerate}
    \item Initiating contact through GitHub pull requests.
    \item Direct email to developers on the Terra-Money repository.
    \item Messages through Discord.
    \item Messages through Telegram.
    \item Scheduling requests to their calendars.
    \item Connection requests through LinkedIn.
    \item Application to job postings on Terra-Money's website.
    \item Invited a TFL member to Terra Rebels GitHub org and assigned co-ownership of the org for 3 weeks. Due to failure to communicate we demoted the individual to a member role.
\end{enumerate}
There have been several TFL developers who have commented and contributed to the implementation and solutions; however, they have acted in good faith independently from the organization.

We will continue in our efforts to communicate with TFL.  We believe this could be a symbiotic relationship.  TFL would have a community of developers maintaining the code base and TFL could provide insight into the 400,000 code commits made over 4 years, including the nuanced workings of the Terra blockchain.  With TFL, improvements to the system would be \textit{orders of magnitude times faster}.  Unfortunately, with the absence of any official announcement, we are operating under the assumption that TFL has already abandoned the chain.  

\subsection{The Community Blockchain}
A Proof-of-Stake (PoS) blockchain can be thought of as a trinity between three entities: validators, community, and developers.  When Luna collapsed, one of the founders of the Luna Blockchain, Do Kwon, asked that developers pledge their support to the Luna 2.0 chain stating, ``Call to action: we encourage Terra developers to signal support \& commit to build on the fork on public channels ASAP.''\footnote{\url{https://twitter.com/stablekwon/status/1526258290316828673}}  Thus, the trinity was broken as the developers abandoned the classic chain and left the community and validators behind.  The community e.g. Luna Classic and UST token holders, also drastically changed as the value of the tokens rapidly lost value, and new members of the community could obtain millions of Luna for only a few USD.  With the absence of developers and an overhaul of the community demographic, the validators are the only members left of the original trinity.  As we speak, validators are slowly leaving the chain with only 90 out of 130 still in the active set.  Even after two months have passed since the depegging event, validation is still broken - delegation and staking on Luna Classic, a PoS blockchain where staking is absolutely fundamental to the operation of the chain, are still disabled.  

However, the Classic ecosystem is not broken.  A new community has rallied behind Luna Classic. The authors of this technical specification are community members that self-organized under a Discord group known as ``Terra Rebels''.  We have no affiliation with Terraform Labs (TFL) nor work under any central organization or established entity at this time.  We, the community, are filling the developer void \textit{impartially and in accordance with community proposals that pass the voting process}.  In the next sections, we describe the proposals passed by the Terra Classic Decentralized Autonomous Organization (DAO) and the associated implementations.  When adopted, a new wave of delegators and validators from the community will stake on Luna Classic, built by community developers.  This will be the true community-owned and operated blockchain.




\section{Proposition 4095 Re-enable Staking/Delegation to Existing Active Validator Set Only}
\noindent This was a text proposal that passed June 21st with a Yes vote of 90.67\% (149M) with a total vote of 165.32M / 321.22M.  The proposal states,
\begin{quote}
This proposal aims to make the following changes: 1) Re-enable Staking/Delegations on Terra Classic exclusively to the current active validator set for a period of 60 days. The ability to create new validators will remain disabled until block height \#8905758 (approximately, 22nd August, 2022). 2) New validators can be created only after block height \#8905758 (approximately, 22nd August, 2022). 
\end{quote}

\subsection{Proposed Solution and Implementation}
\noindent\textbf{Status} - Implemented \\
\textbf{Pull Request} -  \url{https://github.com/terra-money/cosmos-sdk/pull/80}
\begin{lstlisting}
const StakingPowerUpgradeHeight = 7603700
// StakingPowerRevertHeight re-enables the creation of validators after this 
// block height.  This has been computed as approximately August 22, 2022.  68 days from June 15
// With an average of 7 second blocks, there are approximately 8.571 blocks per minute (60/7)
// 8.571 * 60 min * 24 hrs * 68 days = 839,272 blocks
// current block height on June 15 is 8,066,486
// projected block on August 22 is 8,066,486 + 839,272 = 8,905,758
const StakingPowerRevertHeight = 8905758

// DelegatePowerRevertHeight re-enables the ability to delegate stake to existing validators
// This is an approximate block height of adoption.  The exact block height can be agreed upon
// if governance 4095 passes and validators agree to adopt the code patch.
// projected block 5 days from now, June 20 is 
// 8.571 * 60min * 24 hrs * 5 days = 61,711 blocks
// current block height on June 20 is 8,146,938
// projected block on June 25 is 8,146,938 + 61,711 = 8,208,649. 
// This is approximate and will be adjusted if needed
const DelegatePowerRevertHeight = 8208649
\end{lstlisting}
Two block heights are defined, StakingPowerRevertHeight and DelegatePowerRevertHeight.  These are placeholder values, and will be adjusted in consultation with validators.  Once the DelegatePowerRevertHeight is determined, we will compute the StakingPowerRevertHeight to be 60 days after.
\begin{lstlisting}[escapechar=!]
	ctx := sdk.UnwrapSDKContext(goCtx)

	currHeight := ctx.BlockHeight()
	if currHeight > StakingPowerUpgradeHeight && !\colorbox{light-gray}{currHeight < StakingPowerRevertHeight}! {
		return nil, sdkerrors.Wrapf(types.ErrMsgNotSupported, "message type %T is not supported at height %d", msg, currHeight)
	}
\end{lstlisting}
The highlighted change shows that upon reaching the specified revert height, CreateValidator would become functional again.  Similarly, we perform the same revert for Delegation, just at an earlier block,
\begin{lstlisting}[escapechar=!]
func (k msgServer) Delegate(goCtx context.Context, msg *types.MsgDelegate) (*types.MsgDelegateResponse, error) {
	ctx := sdk.UnwrapSDKContext(goCtx)

	currHeight := ctx.BlockHeight()
	if currHeight > StakingPowerUpgradeHeight && !\colorbox{light-gray}{currHeight < DelegatePowerRevertHeight }!{
		return nil, sdkerrors.Wrapf(types.ErrMsgNotSupported, "message type %T is not supported at height %d", msg, currHeight)
	}
\end{lstlisting}
\subsection{Concerns and Potential Pitfalls}
\textit{Concern: The upgrade would cause the chain to halt}.  We have heard the concern that the chain might need to be halted for new code to be adopted.  This should not be the case.  The revert heights protect against consensus failure before the revert height goes into effect.  Thus, validators should be able to upgrade the code at their leisure.  As long as 2/3 of the validators by staking power adopt the code before the first revert height, the chain should not halt. \\

\textit{Concern: Re-enabling staking could result in a BFT attack}.  This attack, where 2/3 of the validators by staking power are malicious actors, is a potential attack for all blockchains.  However, given the imbalance of community-owned Luna and the staked Luna, the attack becomes more plausible when delegation and staking are re-enabled.  We mitigate this attack in two ways.  First, \textit{we only open delegation up to existing validators in the first 60 days}.  An attack in this time period \textit{could only come from the existing validators, who have already demonstrated their commitment to faithfully running the chain}.  While  we acknowledge a non-zero chance of such an attack, we believe it is highly unlikely \cite{swanson2015consensus}.  Further, from a game theoretic perspective, staking would bond the user coins for a period of 21 days, during which time the attack would negatively impact the market price.  Attacking the chain would result in financial loss to the malicious actor, which further strengthens the argument that the attack would be unlikely.  If a wealthy individual or group were to purchase hundreds of millions of dollars to attack the chain, given the market cap of Luna Classic and UST, they would be entitled to it as they have ``paid for it''.  But in this situation, we would roll back to the snapshot before delegation and fork the chain and abandon that version of the chain.  The attacker would suffer catastrophic financial loss.

That does not mean we should not proceed cautiously and take security seriously \cite{postmortem2}.  As a precaution, we will disincentivize such an attack by making the cost to achieve a supermajority stake significantly cost-prohibitive.   We are providing a script to validators that allows them to delegate to their validators on the very first block that the delegation becomes active.  Not only will validators be staking on the first block, we the community will also be delegating our coins equally across a set of trusted validators.  For the validators, we suggest increasing the size of the mempool on the validator nodes and increasing the minimum gas.  We further welcome any additional suggestions to assist with this transition. The prototype script is as follows,
\begin{lstlisting}
#!/bin/bash
var=$(terrad query block)
final=$(echo $var | awk -F '"height":"|","time"' '{print $2}')

while (( $final < $5 ))
do 
	sleep 3
	var=$(terrad query block)
	final=$(echo $var | awk -F '"height":"|","time"' '{print $2}')
done
terrad tx staking delegate $1 $3 --from=$2 --chain-id=rebel-1 --gas=auto --gas-adjustment=1.4  


\end{lstlisting}
This script checks the block height every 3 seconds, and when the targeted block height has been reached, it will delegate a specified amount to a validator account.

\section{Proposition 3568 Tax/Burn 1.2\% all transactions}
\noindent This was a text proposal that passed June 9 with a Yes vote of 83.32\% (140M) with a total vote of 168.13M / 321.22M.  The proposal states,
\begin{quote}
A Tax Burn mechanism is to be implemented on LUNC to reduce the Total Supply. Implement a Tax + Burn mechanism on each buy-sell transaction: 1.2\% burn tax This mechanism should be true until the total supply = 10 billion LUNC, after that, this mechanism is disabled and the total supply can never be changed. this is to be implemented in all transactions and to be suggested officially by the terra team on all social media that all the exchanges should do the same thing until the condition is met. another thing is for the official terra team to share on all social media the official burning address. this will help stop the scams. this is a working process please feel free to share your ideas, this proposal was made because we as a community ear the community and proposal 2975 was with a significant tax, so these numbers are the numbers that the community wants 1.2\%. together we can do this please vote is a lot already being done this is just something that will help speed up things for lunc, and maybe bring back investors and dev and validators to the project. 
\end{quote}

\subsection{Proposed Solution and Implementation}
\noindent\textbf{Status} - Implemented \\
\textbf{Code} -  \url{https://github.com/terra-rebels/classic-core/tree/burntax_via_stability}

The first change we introduce is in the AnteHandler, ante.go.  The AnteHandler is a set of decorators in Tendermint that check transactions on both CheckTx and DeliverTx.  These methods are run on every transaction proposed to the blockchain.
\begin{lstlisting}[escapechar=!]
		cosmosante.NewValidateMemoDecorator(options.AccountKeeper),
		cosmosante.NewConsumeGasForTxSizeDecorator(options.AccountKeeper),
		cosmosante.NewDeductFeeDecorator(options.AccountKeeper, options.BankKeeper, options.FeegrantKeeper),
    		!\colorbox{light-gray}{NewBurnTaxFeeDecorator}!(options.TreasuryKeeper, options.BankKeeper),           // burn tax proceeds
		cosmosante.NewSetPubKeyDecorator(options.AccountKeeper), // SetPubKeyDecorator must be called before all signature verification decorators
		cosmosante.NewValidateSigCountDecorator(options.AccountKeeper),
		cosmosante.NewSigGasConsumeDecorator(options.AccountKeeper, sigGasConsumer),
\end{lstlisting}
Here we adopt and define a new decorator called NewBurnTaxFeeDecorator.  This is nearly identical to the NewTaxFeeDecorator introduced by TFL for implementing the stability tax on UST.  What is different is that we pass in the BankKeeper module to burn the tax immediately after subtracting the tax from the sender, i.e. immediately after the cosmosante.NewDeductFeeDecorator.

Here we show the burntax.go file which is the implementation of the new decorator.  Again, one can easily verify the identical nature of this file with tax.go (the TFL stability tax decorator).

\begin{lstlisting}[escapechar=!]
package ante

import (
	"fmt"

	sdk "github.com/cosmos/cosmos-sdk/types"
	sdkerrors "github.com/cosmos/cosmos-sdk/types/errors"
	"github.com/cosmos/cosmos-sdk/x/auth/types"
)

// BurnTaxFeeDecorator will immediately burn the collected Tax
type BurnTaxFeeDecorator struct {
	treasuryKeeper TreasuryKeeper
	!\colorbox{light-gray}{bankKeeper BankKeeper}!
}

// NewBurnTaxFeeDecorator returns new tax fee decorator instance
func NewBurnTaxFeeDecorator(treasuryKeeper TreasuryKeeper, bankKeeper BankKeeper) BurnTaxFeeDecorator {
	return BurnTaxFeeDecorator{
		treasuryKeeper: treasuryKeeper,
		!\colorbox{light-gray}{bankKeeper: bankKeeper,}!
	}
}
\end{lstlisting}
As described before, the only difference is that we bring in the Bank module to handle burning the tax on every eligible transaction.

The AnteHandler then computes the taxes in the exact same way as the stability tax.  However, here instead of just recording the tax proceeds as the stability tax used to, we instead send the coins from the FeeCollector module that collected the gas plus tax, to the BurnModule that handles reducing the total supply.
\begin{lstlisting}[escapechar=@]
// AnteHandle handles msg tax fee checking
func (btfd BurnTaxFeeDecorator) AnteHandle(ctx sdk.Context, tx sdk.Tx, simulate bool, next sdk.AnteHandler) (newCtx sdk.Context, err error) {
	feeTx, ok := tx.(sdk.FeeTx)
	if !ok {
		return ctx, sdkerrors.Wrap(sdkerrors.ErrTxDecode, "Tx must be a FeeTx")
	}

	msgs := feeTx.GetMsgs()

	//At this point we have already run the DeductFees AnteHandler and taken the fees from the sending account
	//Now we remove the taxes from the gas reward and immediately burn it

	if !simulate {
		// Compute taxes again.  Slightly redundant
				taxes := FilterMsgAndComputeTax(ctx, btfd.treasuryKeeper, msgs...)

		// Record tax proceeds
		if !taxes.IsZero() {
				@\colorbox{light-gray}{btfd.bankKeeper.SendCoinsFromModuleToModule}@(ctx, @\colorbox{light-gray}{types.FeeCollectorName, treasury.BurnModuleName}@, @\colorbox{light-gray}{taxes}@)
			if err != nil {
				return ctx, sdkerrors.Wrapf(sdkerrors.ErrInsufficientFunds, err.Error())
			}

\end{lstlisting}

In order for us to actually send tokens from the FeeCollector to the BurnModule, we do need to expose the SendCoinsFromModuleToModule endpoint to the AnteHandler, so we tell the handler that we can expect to access this function via the expected\_keeper.go file.  This code gives the AnteHandler access so that it can burn the coins directly.
\begin{lstlisting}[escapechar=!]
// BankKeeper defines the contract needed for supply related APIs (noalias)
type BankKeeper interface {
	SendCoinsFromAccountToModule(ctx sdk.Context, senderAddr sdk.AccAddress, recipientModule string, amt sdk.Coins) error
	!\colorbox{light-gray}{SendCoinsFromModuleToModule}!(ctx sdk.Context, senderModule string, recipientModule string, amt sdk.Coins) error
}
\end{lstlisting}

Now the actual tax percentage amount is obtained via the existing treasury parameter TaxRate.  This tax rate is modulated by the tax policy and treasury module as defined here \cite{treasury}.  The tk.GetTaxRate method queries this parameter.

\begin{lstlisting}[escapechar=!]
// computes the stability tax according to tax-rate and tax-cap
func computeTax(ctx sdk.Context, tk TreasuryKeeper, principal sdk.Coins) sdk.Coins {
	taxRate := !\colorbox{light-gray}{tk.GetTaxRate(ctx)}!

\end{lstlisting}

Finally, there was logic in both the existing tax AnteHandler and in the treasury module to not tax Luna.  We removed the limitation of taxing Luna so that the burn would apply. 

\begin{lstlisting}[escapechar=!]
	for _, coin := range principal {
		//Originally only a stability tax on UST.  Changed to tax Luna as well.
		//if !\colorbox{light-gray}{coin.Denom == core.MicroLunaDenom}! || coin.Denom == sdk.DefaultBondDenom {
		if coin.Denom == sdk.DefaultBondDenom {
			continue
		}
\end{lstlisting}

treasury/keeper/keeper.go
\begin{lstlisting}[escapechar=!]
// GetTaxCap gets the tax cap denominated in integer units of the reference {denom}
func (k Keeper) GetTaxCap(ctx sdk.Context, denom string) sdk.Int {
	// allow tax cap for uluna
		//if !\colorbox{light-gray}{denom == core.MicroLunaDenom}! {
	//	return sdk.ZeroInt()
	//}
\end{lstlisting}

The following delineates the exact parameter changes that need to be passed by governance in order to activate and properly burn the taxed tokens with a 1.2\% tax.
\begin{lstlisting}
{
  "title": "Param Change Policy",
  "description": "Update tax policy",
  "changes": [
    {
      "subspace": "treasury",
      "key": "TaxPolicy",
      "value": {"rate_min": "0.012", "rate_max": "0.012", "cap":{"denom": "usdr", "amount": "10000000" }, "change_rate_max": "0.0"}
    },
    {
      "subspace": "treasury",
      "key": "RewardPolicy",
      "value": {"rate_min": "1.0", "rate_max": "1.0", "cap":{"denom": "unused", "amount": "0" }, "change_rate_max": "0.0"}
    }
  ],
}
\end{lstlisting}
The rate\_min and rate\_max parameters should be set to 0.012 to achieve the 1.2\%.  The tax will be clamped to exactly this number.  There is a cap on the total tax allowed in usdr units.  This can be set arbitrarily high.  Setting the change\_rate\_max to 0.0 prevents the tax rate from changing over epochs (approximately a week) of the blockchain.  An epoch must pass for the tax rate to be enacted.

The reward policy rate\_min and rate\_max should be set to 1.0, with a change\_rate\_max set to 0.0.  This requires some explanation.  At the end of an epoch, the treasury calculates the total number of tokens that have been burned in that epoch.  \textit{It then immediately mints that exact amount} and distributes these as rewards to the validators and community pool as part of the old seignorage model.  The reward policy specifies how much of this should be burned, and how much should be distributed.  1.0 or 100\% indicates that all of the newly minted Luna should immediately be burned and not distributed.  

\textit{It is imperative that these parameter changes be made simultaneously in the same proposal.  Otherwise, the tax could be exploited for rewards, and not burned.}

\subsection{Concerns and Potential Pitfalls}
\textit{Concern: The tax is bad for long-term growth}.  \\
Whether or not this tax is a positive or negative catalyst for the Luna Classic ecosystem remains to be seen.  We remain impartial and develop in accordance with community proposals that pass the voting process.   While we acknowledge that taxing transactions will likely reduce the activity on the chain, we have structured the implementation to be flexible as the Luna Classic landscape evolves.  The tax can be changed via parameter proposal at any time, and we could even take advantage of the variable tax rate to adjust after every epoch.  Through a democratic process, the Luna Classic community can adapt to maintain healthy long-term growth.\\

\textit{Concern: The tax will break connections to certain dApps and exchanges}.\\
This is likely true.  We have already seen issues arise internally with tools like Terra Station, Terra Faucet, and Terra's chrome extension, which are designed to probe the blockchain using a dummy transaction of 1 Luna to compute the gas fees.  When increasing the amount being transacted, the gas fees are not recomputed, and the transaction will fail.  

A large coordinated marketing and publicity campaign will need to occur when governance of the parameter proposal indicates that this will pass.  Developers and third parties will need to be contacted to upgrade their logic for interacting with the Classic blockchain.  

However, we see a silver lining.  Our code change is based upon the UST stability tax that was \textit{already in place and active up until February 2022.} Governance proposal 172 reduced this tax rate to 0.  This means the developers and exchanges that were utilizing the UST stablecoin already have this logic, and it would be a matter of re-enabling the same logic for Luna Classic.

\section{Proposition 4080 Distribution 50\% Transaction Fees to Community Pool}
\noindent This was a parameter proposal that passed June 15th with a Yes vote of 94.79\% (181M) with a total vote of 191.23M / 321.22M.  The proposal states,
\begin{quote}
Distribute 50\% transaction fees to the community pool (35\% to be burned via monthly community pool proposals; 10\% airdropped to ecosystem devs, 5\% retained for core Terra Classic development) and increase 'Base Proposer' and ‘Bonus Proposer' reward from 0.01 and 0.04 to 0.03 and 0.12 respectively. 
\end{quote}
\begin{lstlisting}
{
  "subspace": "distribution",
  "key": "communitytax",
  "value": "0.500000000000000000"
}
{
  "subspace": "distribution",
  "key": "baseproposerreward",
  "value": "0.030000000000000000"
}
{
  "subspace": "distribution",
  "key": "bonusproposerreward",
  "value": "0.120000000000000000"
}
\end{lstlisting}

\subsection{Implemented Solution}
\noindent\textbf{Status} - Active \\
This was a parameter change proposal and so the parameters were automatically implemented by the governance module.

\subsection{Concerns and Potential Pitfalls}
\textit{Concern: Who will initiate the burn proposal every month?} \\
At this time, it is unclear who will initiate the burn proposal every month.  While the authors are in communication with the creator of this proposal, we are unaware of any formal schedule or responsibilities put in place for the community burn.

\textit{Concern: What prevents the abuse of the community funds?} \\
The distribution of the community pool is defined by a distribution governance proposal.  The community and validators will have to vote for the distribution of these funds.

\section{Proposition 1299 (Re)Enable IBC}
\noindent This was a parameter proposal that passed with a Yes vote of 95.14\% (150M) with a total vote of 158.15M / 321.22M.  The proposal states,
\begin{quote}
Terra validators disabled IBC as a stop-gap solution to preventing Impermanent Loss on UST and LUNA pools on Osmosis and other IBC DEXs. Unfortunately, this also prevents UST and LUNA from transferring between chains. Currently, about 154.7M UST is stuck in Osmosis alone. This proposal re-enables and unlocks the transfer of UST and LUNA between chains.
\end{quote}

\begin{lstlisting}
Changes
    {
      "subspace": "transfer",
      "key": "SendEnabled",
      "value": "true"
    }
    {
      "subspace": "transfer",
      "key": "ReceiveEnabled",
      "value": "true"
    }
\end{lstlisting}

\subsection{Proposed Solution and Implementation}
\noindent\textbf{Status} - Not Implemented \\
While this parameter proposal passed, IBC was disabled in code, not in the parameter space.  The technical code for re-enabling IBC is trivial, but is not implemented at this time.  See the concerns and potential pitfalls.
\subsection{Concerns and Potential Pitfalls}
During the UST depeg, there was significant concern regarding IBC vulnerabilities.  We have reached out to 2 Cosmos developers, one that closed the channel in the first place, and one that warned against re-opening early when UST was depegging.   We plan to talk to inter-chain teams, like Osmosis and Juno, and consult the Cosmos developers who were involved in the first place.  Further research is warranted to ensure the re-opening of IBC is secure.

\section{Rebel-1 TestNet}
On June 24, 2022, ``Rebel-1'' TestNet went live with the proposed code changes for propositions 3568 and 4095.  In the span of 48 hours, the TestNet was supported by 7 community validators donating their time and hardware.  Now, after a week, the TestNet has 11 validators and new members are being onboarded daily.  The TestNet is able to be accessed through the existing Terra Station desktop app and chrome extension.  The distribution of funds is automatic through a Faucet NodeJS application.

For our code changes, the 4095 code was tested to see if our logic would disable delegation, then enable delegation after a specified block.  This was successful.  Our 3568 test included whether the correct tax was being computed with on-chain transactions.  We also tested the governance parameter proposals that would need to be set in order for the burn tax to be executed as designed.  Our tests have concluded that the code is working as expected.  Integration tests on Columbus-5 is ongoing.

\section{Conclusion and Future Directions}
In future directions, we believe the algorithmic stablecoin connection between Luna Classic to UST is an important, useful, and differentiating factor of the Terra ecosystem.  This topic is beyond the scope of emergency measures that need to be taken immediately and will be addressed more thoroughly in a different document.  

In conclusion, we have presented our technical specifications on how the passed governance proposals can be implemented.  We publicly audit our code in order to build trust from the community and validators.  We also outline our concerns about the various proposals for full transparency.  Finally, we ask the larger community for assistance in vetting and implementing the code base.  \textit{We are adopting a legacy code base that was not developed by us, nor are being given guidance by the creators.}  We ask that those who have the knowledge or resources contact us to help rebuild the Terra Classic ecosystem.

\textit{Financial Disclaimer} - No author is being financially compensated for their assistance in emergency management of the Luna Classic blockchain.  Several authors have disclosed personal financial interest in the form of LUNC and UST holdings.

 \bibliographystyle{elsarticle-num} 
 \bibliography{cas-refs}
 
 \newpage
 \section{Appendix}
 Following the release of our technical specification, we performed unit and integration tests on all the code changes, and added a new protection mechanisms to the staking and delegation code.  We outline the results of the tests and code changes in the following appendix.
 \section{Proposition 4095 Test Results}
\noindent\textbf{Status} - Success \\

\noindent Rebel-1 TestNet was downgraded to v0.5.20 (v20) on July 8th, 2022.  This version is the current software running on MainNet, columbus-5.  A new genesis was created that started the TestNet on block 7561000, approximately 72 hours before the de-pegging event.  As per the security patch by TFL, staking and delegation was disabled at block 7603700.  We were able to recruit 8 validators before the security block, 2 of which became inactive over time.  Thus, only 6 validators were active participants in the upgrade test process.  Each validator was delegated approximately 10,000 Lunc (+/- 1000) giving them approximately equal voting power.  

After the security patch became active on July 13th, 2022, v0.5.21-testnet (v21), was distributed to validators.  The validators then upgraded asynchronously between July 16th, 2022.  Four out of the six validators confirmed their upgrades to v21.  Both v20 and v21 were able to co-exist on the network simultaneously during this upgrade.  We then reached the DelegatePowerRevertHeight at block 7684490.  
 
\begin{figure}[ht]
 \centering
{\includegraphics[width=9.05cm]{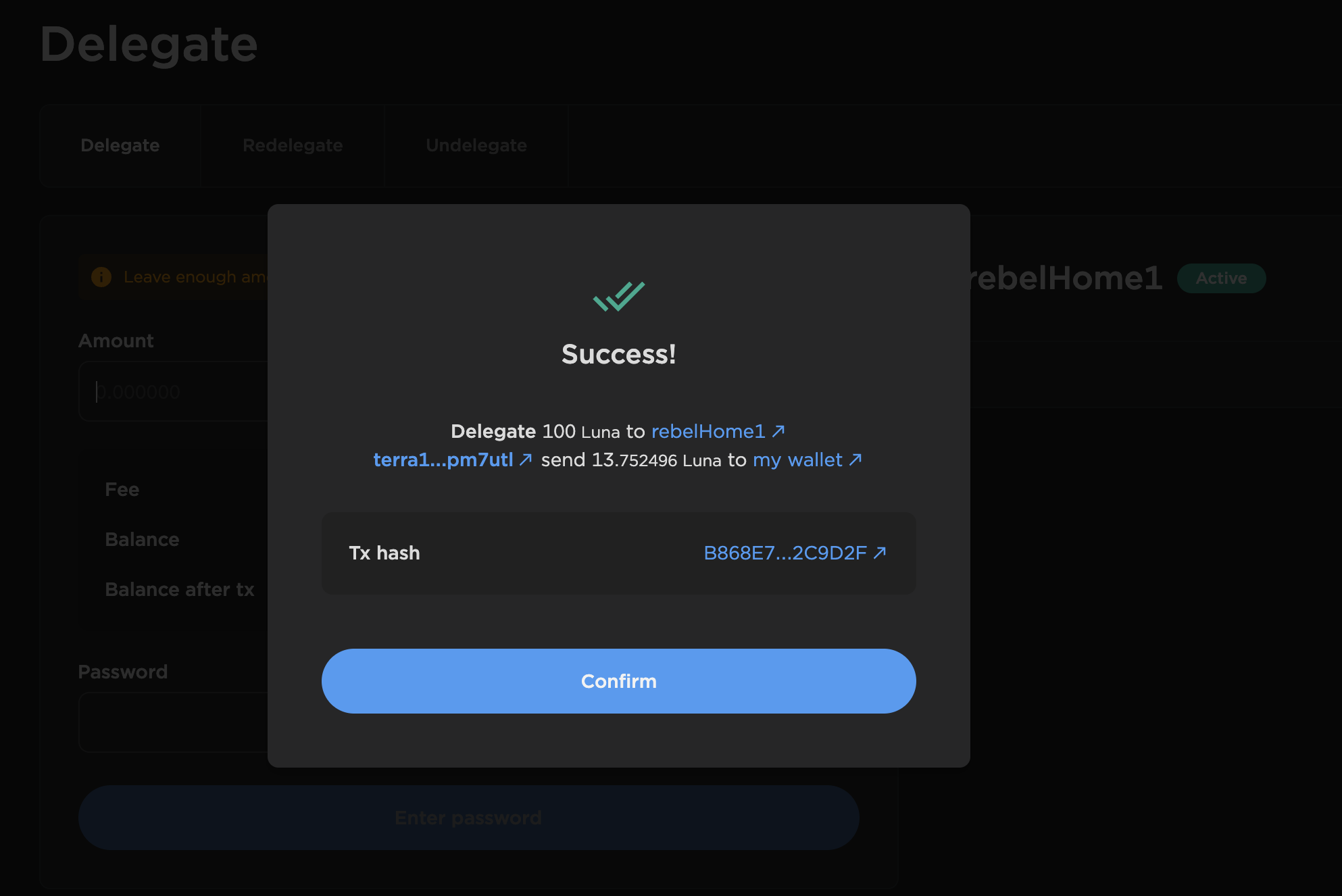}}
  \caption{Delegation was successful after the DelegatePowerRevertHeight at block 7684490 on TestNet.  This image was capture via terra-station.}
\label{fig:delegation}
\end{figure}

Our script issued a delegate transaction command and the chain halted at block height 7684492, due to the fact that approximately 64\% of the voting power had upgraded (less than 2/3s).  The other validators had consensus failures and were required to be restored from a prior snapshot.  The chain was halted for approximately 15 minutes as the other validators were upgraded to v21 and then resumed.  Delegation via the terrad command line and terra-station was successful, see Figure \ref{fig:delegation}. 

We previously had a concern where the chain could be attacked if 2/3 of the validators by staking power are malicious actors.   This attack was noted as a concern by members of the community, including existing validators.  While our previous recommendations still stand (see Section 2.2), we implemented a stronger security measure directly in the code base.  This security measure checks every delegation transaction and computes whether or not it will increase the validator's voting power over a certain limit.  In this case, we determined that no validator should achieve more than 25\% of the voting power.  If the limit is exceeded, the command will fail.  This security measure will be in place until the ProtectPowerHeight block is passed, which we propose to be 60 days.
 \begin{lstlisting}
 
// If Delegations are allowed, but we are in a vulnerable state below ProtectPowerHeight, limit validator power
if currHeight >= DelegatePowerRevertHeight && currHeight < ProtectPowerHeight {
        // Get the Total Consensus Power of all Validators
        lastPower := k.Keeper.GetLastTotalPower(ctx)
        ctx.Logger().Info(fmt.Sprintf("lastPower is %s", lastPower))

        // Get the selected Validator's voting power
        validatorLastPower := k.Keeper.GetLastValidatorPower(ctx, valAddr)
        ctx.Logger().Info(fmt.Sprintf("lastPower of Validator is %d", validatorLastPower))

        // Compute what the Validator's new power would be if this Delegation goes through
        validatorNewPower := int64(validatorLastPower) + sdk.TokensToConsensusPower(msg.Amount.Amount, k.Keeper.PowerReduction(ctx))

        // Compute what the Total Consensus Power would be if this Delegation goes through
        newTotalPower := lastPower.Int64() + sdk.TokensToConsensusPower(msg.Amount.Amount, k.Keeper.PowerReduction(ctx))
        ctx.Logger().Info(fmt.Sprintf("newPower of Validator would be %d", validatorNewPower))

        // Compute what the new Validator voting power would be in relation to the new total power
        validatorIncreasedDelegationPercent := float32(validatorNewPower) / float32(newTotalPower)

        // If Delegations are allowed, and the Delegation would have increased the Validator to over 25% of the staking power, do not allow the Delegation to proceed
        if validatorIncreasedDelegationPercent > 0.25 {
                return nil, sdkerrors.Wrapf(types.ErrMsgNotSupported, "message type %T is over the allowed limit at height %d", msg, currHeight)
        }
}
  \end{lstlisting}
  
  Finally, the last test for Proposition 4095 is the creation of new validators.  On July 18th, 2022, the TestNet hit the revert staking block and new validators were successfully added to the network.
  
  \section{Proposition 3568 Test Results}
\noindent\textbf{Status} - Success \\ 

\noindent Include in v0.5.21 is the burn tax code implemented via a new burn AnteHandler method.  During synchronization of v21 on columbus-5, we ran into an error where a smart contract on chain was querying the TaxCap.  Thus, we decided to enable all of the tax burn code after a specific block height, TaxPowerUpgradeHeight.  However, just because the tax block height passes, does not mean the taxes are active.  Instead, the tax must be initiated by a parameter change governance proposal as described above.  On TestNet, we successfully initiated and passed the governance proposal for a 1.2\% tax.  On block height 7684490, all transactions on the TestNet were successfully taxed.

We completed unit testing for all possible messages that are tax eligbile.  This includes, MsgSend, MsgMultiSend, MsgSwapSend, MsgIntantiateContract, MsgExecuteContract, and MsgExec.  Note that all staking and governance related transactions do not incur tax.  Here we show the results of the unit tests for all of these transactions related to the Luna Classic tax highlighted in gray.

\begin{lstlisting}[escapechar=!]
--- PASS: TestAnteTestSuite/TestConsumeSignatureVerificationGas (0.00s)
--- PASS: TestAnteTestSuite/!\colorbox{light-gray}{TestEnsureBurnTaxModule}! (0.01s)
--- PASS: TestAnteTestSuite/TestEnsureMempoolFeesExec (0.01s)
--- PASS: TestAnteTestSuite/!\colorbox{light-gray}{TestEnsureMempoolFeesExecLunaTax}! (0.01s)
--- PASS: TestAnteTestSuite/TestEnsureMempoolFeesExecuteContract (0.01s)
--- PASS: TestAnteTestSuite/!\colorbox{light-gray}{TestEnsureMempoolFeesExecuteContractLunaTax}! (0.01s)
--- PASS: TestAnteTestSuite/TestEnsureMempoolFeesGas (0.01s)
--- PASS: TestAnteTestSuite/TestEnsureMempoolFeesInstantiateContract (0.01s)
--- PASS: TestAnteTestSuite/!\colorbox{light-gray}{TestEnsureMempoolFeesInstantiateContractLunaTax}! (0.01s)
--- PASS: TestAnteTestSuite/TestEnsureMempoolFeesMultiSend (0.01s)
--- PASS: TestAnteTestSuite/!\colorbox{light-gray}{TestEnsureMempoolFeesMultiSendLunaTax}! (0.01s)
--- PASS: TestAnteTestSuite/TestEnsureMempoolFeesSend (0.01s)
--- PASS: TestAnteTestSuite/!\colorbox{light-gray}{TestEnsureMempoolFeesSendLunaTax}! (0.01s)
--- PASS: TestAnteTestSuite/TestEnsureMempoolFeesSwapSend (0.01s)
--- PASS: TestAnteTestSuite/!\colorbox{light-gray}{TestEnsureMempoolFeesSwapSendLunaTax}! (0.01s)
--- PASS: TestAnteTestSuite/TestIncrementSequenceDecorator (0.01s)
--- PASS: TestAnteTestSuite/TestOracleSpamming (0.01s)
--- PASS: TestAnteTestSuite/TestSetPubKey (0.01s)
--- PASS: TestAnteTestSuite/TestSigIntegration (0.01s)
--- PASS: TestAnteTestSuite/TestSigVerification (0.01s)
--- PASS: TestAnteTestSuite/TestSigVerification_ExplicitAmino (0.01s)
\end{lstlisting}
  
\subsection{Recommendation}
While we are confident that the tax code is working correctly on-chain, it is not clear that external applications and exchanges would be computing the appropriate fees.  \textbf{At this time it is our recommendation that v21 be adopted by validators, but the tax code should not be activated by parameter proposal until compliance by important internal and external entities is confirmed e.g. Terrastation, Binance, Kucoin, Anchor, Terraswap, etc.}  It would be imperative that v21 be available and distributed to CEXs, DEXs, and other dApp developers active on Luna Classic.  Without the appropriate computation of the fees required, transactions will fail with insufficient gas.  
  
  The proper way to compute the necessary gas fee is the following equation,
\begin{equation}
    newFee = min(taxRate \times amount, taxCap) + gas
\end{equation}
where the taxRate is to be set to 1.2\%, the amount is the total amount being sent, taxCap is the maximum tax allowable, and gas is the old gas fees that are estimated.  In the current passed governance, the taxCap should be set appropriately high so that this is greater than 1.2\% of the total supply.

Smart contracts can compute the estimated gas fees using the sample code shown below.  This is similar to the code used by TerraSwap and Astroport.
 \begin{lstlisting}
 pub fn deduct_tax(&self, querier: &QuerierWrapper) -> StdResult<Coin> {
    let amount = self.amount;
    if let AssetInfo::NativeToken { denom } = &self.info {
        Ok(Coin {
            denom: denom.to_string(),
            amount: amount.checked_sub(self.compute_tax(querier)?)?,
        })
    } else {
        Err(StdError::generic_err("cannot deduct tax from token asset"))
    }
}

 \end{lstlisting}





\end{document}